# The high osmotic pressure in a lens fiber as a driving force in the development of senile cortical cataract

Jicun Wang-Michelitsch[1]*, Thomas M Michelitsch[2]

[1]Department of Medicine, Addenbrooke's Hospital, University of Cambridge, UK (work address until 2007)

[2]Institut Jean le Rond d'Alembert (Paris 6), CNRS UMR 7190 Paris, France

**Abstract**

In lens cataract, the clouding change in lens leads to a decline of transparency of part of the lens. There are three types of senile cataract: cortical cataract, nuclear cataract, and posterior/anterior sub-capsular cataract. The most common senile cataract is cortical cataract. For understanding cortical cataract, the pathology and the causing factors in cortical cataract are analyzed. The clouding change in senile cortical cataract begins from the edge of the lens and develops progressively to lens centre. The pathology of clouding change in cortical cataract is characterized by disruption of some lens fibers, swelling of some other fibers, and deposition of water between fibers. Based on the property of a lens fiber, we propose here a hypothesis on the mechanism of development of senile cortical cataract. **I.** Cortical cataract is triggered by disruption of a cortical lens fiber as a result of injury. The disrupted fiber will release water and crystallin proteins. **II.** Neighbor fibers can absorb this water due to high intracellular osmotic pressure (IOP) and become swollen. Swelling makes a fiber be stiff and have increased risk to disrupt when it is curved during the accommodation of the lens. These fibers will release water again when they disrupt, and the water will make more local fibers swelling. In this way, the local fibers become swollen and then disrupt successively. **III.** The successive swellings and disruptions of local fibers result in enlargement of a clouding change in lens. Since the fibers on the outer part of lens cortex have higher risk to be injured than that in the inner part, a clouding change starts from the edge of the lens. In conclusion, the progressive development of senile cortical cataract is a result of successive swellings and disruptions of local lens fibers, and this is driven by the high IOP in a lens fiber.

**Keywords**

Lens cataract is a disease of the lens that affects most old people. The clouding change in cataract leads to an irreversible decline of lens transparency. There are three types of senile cataract: nuclear cataract, cortical cataract, and posterior/anterior sub-capsular cataract. Among them, cortical cataract has the highest incidence in old people. The clouding change in senile cortical cataract begins from the edge of lens cortex and develops progressively to the inner part of cortex, appearing as aging of the lens (Van den Brom, 1990). On interpreting aging, we proposed a novel theory, the Misrepair-accumulation theory (Wang, 2009). Interestingly, cortical cataract cannot be explained by this theory. The body of lens cortex and lens centre is a non-living structure, but Misrepair mechanism applies only to living-structures. After analyzing the property of lens fibers, we found out that the high intracellular osmotic pressure (IOP) in a lens fiber might play an important role in the progressive development of cortical cataract. On this basis, we propose in the present paper a hypothesis on the mechanism of development of cortical cataract. Our discussion tackles the following issues:

I. Characteristic of cortical cataract: progressive degeneration of lens fibers

II. Effect of high intracellular osmotic pressure (IOP) on a lens fiber

III. A hypothesis on the development of cortical cataract

IV. The high IOP of a lens fiber as a driving force in the development of cortical cataract

## I. Characteristic of cortical cataract: progressive degeneration of lens fibers

The lens is a transparent and curvable structure in the anterior part of an eye, and it is important in refracting and focusing light on retina. Curvature of the lens can be regulated by ciliary muscles for accommodation.  A lens consists of two parts on structure: lens envelope and lens fibers. Lens capsule is the outer layer of lens envelope and it is made of basement membrane. Beneath the anterior capsule is a layer of lens epithelium. Lens epithelium is the unique living tissue in the lens. In the body of lens fibers, the outer part is lens cortex and the central part is lens center. Lens fiber cells are a type of epithelial cells. A new cell that is produced by lens epithelium needs to differentiate on the lens equator to become a lens fiber. New fibers distribute parallelly on the surface of the lens. Old fibers are being pushed by new fibers into lens centre, during which the fibers lose gradually their water and nuclei. The lens fibers in cortex contain water, but those in lens centre have lost water completely.  It is the parallel arrangement of lens fibers that makes the lens transparent. Continuous production of new lens fibers is essential for maintaining the transparency of the lens.

An alteration of arrangement of lens fibers will result in degeneration of part of the lens, which appears as a clouding change. In senile cortical cataract, degeneration of the lens is pathologically characterized by disruption of some fibers, swelling of some other fibers, and deposition of water between fibers. Interestingly, degeneration of the lens begins always from the edge of lens cortex, which appears as a radial-like or wheel-like clouding change. With time, the clouding change develops progressively to the inner part of lens cortex. One may ask, - What is the element that causes the disruption and the swelling of lens fibers in cortical cataract? - Why does degeneration of the lens begin from the outer part of lens cortex? In our view, the property of a lens fiber, especially the high intracellular osmotic pressure (IOP) of a lens fiber plays an important role in the progressiveness of clouding change in senile cortical cataract.

## II.     Effect of high intracellular osmotic pressure (IOP) on a lens fiber

A lens fiber is a fiber-like non-living "cell" with lipid membrane. In our view, some properties of a lens fiber are similar to that of a keratinocyte in the skin. Keratinocytes are the outer layers of epithelial cells in skin, but they have lost the functionality as cells. However, such a non-living cell has great potential of absorbing water from environment due to the high concentrated keratins inside the cell. Namely, the keratins make a high intracellular osmotic pressure (IOP) for a keratinocyte. The IOP in a keratinocyte is similar to the colloid osmotic pressure in bloodstream, which is made by white proteins. The ability of absorbing water is important for the functionality of keratinocytes. Filled with water, these cells are "full" in shape and able to adhere densely to each other. Several layers of keratinocytes build up a dense and thick non-living wall in skin, for preventing loss of water and invasion of external substances. When new keratinocytes are produced in deeper layers of epithelium, old keratinocytes are pushed up to skin surface. Old keratinocytes lose gradually water and finally drop off from the skin.

Similarly, a lens fiber in lens cortex is a dead cell that contains water and crystallin proteins. In a lens fiber, there are three types of crystallin proteins: crystallin α, crystallin β, and crystallin γ. A regular organization of these three types of crystallin proteins is the basis for the transparency of a lens fiber. However, the high concentrated crystallin proteins give a lens fiber not only transparency but also a high IOP. Due to the high IOP, a lens fiber can absorb water from environment. Filled with water, the fibers can be densely packed, which is important for the tight arrangement of lens fibers. When a new fiber cell is still alive, it is able to control the entrance of water. However, when the fiber is dead, it will lose this ability. A dead fiber can become swollen and stiff when it absorbs too much water by IOP.

## III.    A hypothesis on the development of cortical cataract

It is known that an injury on lens fibers may result in a clouding change of the lens. Except accidents, three factors may cause small but repeated injuries to lens fibers: UV-radiation, high temperature, and the accommodation of the lens. UV-radiation can damage directly lens fibers, and high temperature can make fibers dry and stiff. Repeated adjustments of lens fibers may cause disruption of some fibers occasionally. Stiffness of a fiber, which is caused by high temperature or UV-radiation, will increase the risk of disruption of the fiber when it is curved during lens accommodations. Among all of the lens fibers, those ones that are in the outer part of cortex have the highest risk to be injured by UV-radiation or by heating. Thus, the fibers on the surface of cortex have the highest risk to disrupt during lens accommodations.

In our view, disruption of a fiber in lens cortex is a trigger for the progressive degeneration of lens cortex, because it promotes a series of changes to local lens fibers. When a cortical fiber disrupts, water and crystallin proteins will be released and depose between fibers. The released proteins may degenerate due to change of environment, and aggregate into globing bodies, called Morgagnian-globules. Due to high IOP, neighbor fibers will absorb the water and become swollen gradually. A swollen fiber is stiff and difficult to curve, thus it may disrupt during lens accommodations. Additionally, disruption of a fiber produces a "water gap" between fibers and this may increase as well the physical load to neighbor fibers during adjusting of the lens. The disrupted fibers will release water again, which can make more neighbor fibers swelling. In this way, swellings and disruptions take place to more and more local fibers successively (Figure 1A, 1B, and 1C). In lens nucleus, disruption of a central fiber may not cause swelling of its neighbor fibers, since a nuclear fiber is too dry to release water.

Swelling or disruption of a fiber may appear as a tiny clouding change; and the subsequent swellings and disruptions of neighbor fibers will result in enlargement of the clouding change. Such a process can take place in the same time in different areas of lens cortex, resulting in development of multiple clouding changes. This gives a radial-like or/and a wheel–like distribution of clouding changes in **early stage of cataract**. Degeneration of lens fibers start from the edge of lens cortex and develop into deep cortex by two reasons: **A**. the lens fibers in the outer part of lens cortex have higher risk to be injured than that in the inner part; and **B**. the disrupted fibers will be pushed gradually into lens centre by new fibers.

In early stage of cortical cataract, the procedure of swelling-disruption of a lens fiber is slow, because the released water from a broken fiber is very limited. However, driven by the high IOP, disruption of a swollen lens fiber, releasing of water, and swelling of neighbor fibers compose a viscous circle, which makes local fibers disrupt successively without stop (Figure 2). With time, the number of disrupted fibers increases and the amount of released water augments, thus the procedure of swelling-disruption of a lens fiber is accelerated. As a result, the growth of a clouding change is slow in the beginning but becomes rapider with time. Finally, the lens fibers in deeper layers of cortex will be affected. The swelling of a great number of fibers makes the whole lens swollen and stiff in **swollen stage of cataract**. After disruption of many swollen fibers, the degree of swelling of the lens is reduced and the cataract becomes mature (**Mature stage of cataract**) (Figure 1D). When most of the lens fibers disrupt, lens liquefaction will be observed (**Hypermature stage of cataract**). The liquefied lens is soft, and the lens capsule is fragile in this stage. If the capsule is broken, dissolution of the lens and releasing of crystallin proteins may provoke severe complications in the eye. It may take more than 20 years for a cataract to develop from a radial-like clouding to a complete clouding of the lens.

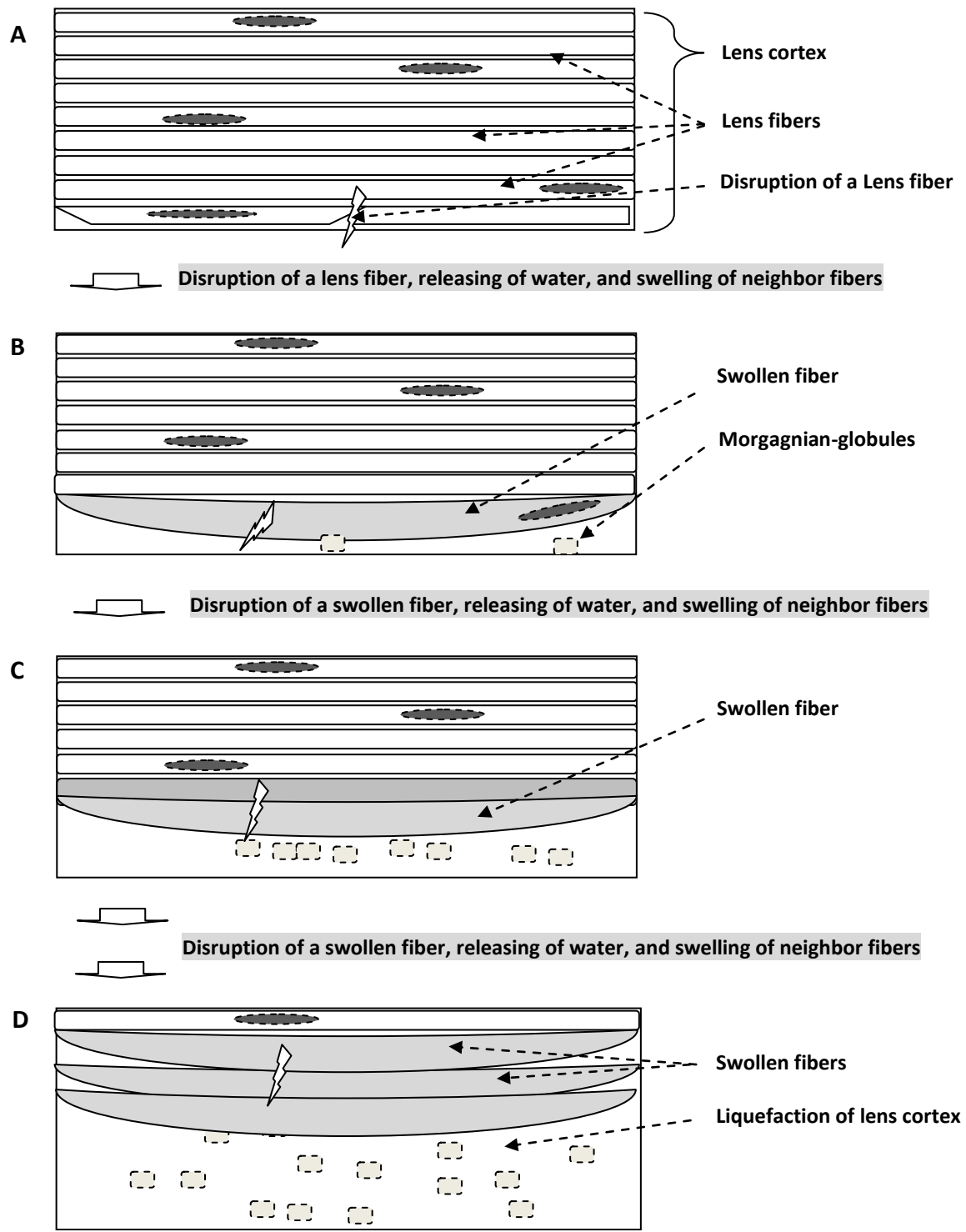

**Figure 1. Successive swellings and disruptions of lens fibers in the development of cortical cataract**

Development of senile cortical cataract is triggered by disruption of a lens fiber in lens cortex. The disrupted fiber will release water and crystallin proteins, and the crystalline proteins may degenerate and aggregate into globing bodies (**Morgagnian-globules**) (**A-B**). Due to high intracellular osmotic pressure (IOP), neighbor fibers will absorb the water and become swollen (**B**). Swelling makes a fiber be stiff and have increased risk to disrupt when it is curved during lens accommodations. The swollen fibers will release water again when they disrupt (**B-C**). In this way, more and more local fibers become swollen and then disrupt successively (**C-D**). When most of the cortical fibers disrupt, lens cortex is liquefied (**D**).

Senile nuclear cataract is rare, and it occurs mainly to the individuals who have had prolonged exposure to UV-light and the individuals who have had problems of high myopia. Differently from cortical cataract, the lens change in nuclear cataract is brown but not grey. One reason for this difference is that the fibers in lens nucleus are dry and disrupted fibers cannot release water and cause swelling of neighbor fibers. The clouding change in nuclear cataract starts from lens centre and may develop gradually to lens cortex; however the process is rather slow. In our view, the development of nuclear cataract is simply a result of accumulation of disruptions of the fibers in lens nucleus. Since disruption of a nuclear fiber does not affect its neighbor fibers by an IOP, the rate of accumulation of injuries of fibers is slow. Such fiber injuries in deep lens may be often caused by severe UV light-exposure or over-adjustments of the lens in high myopia.

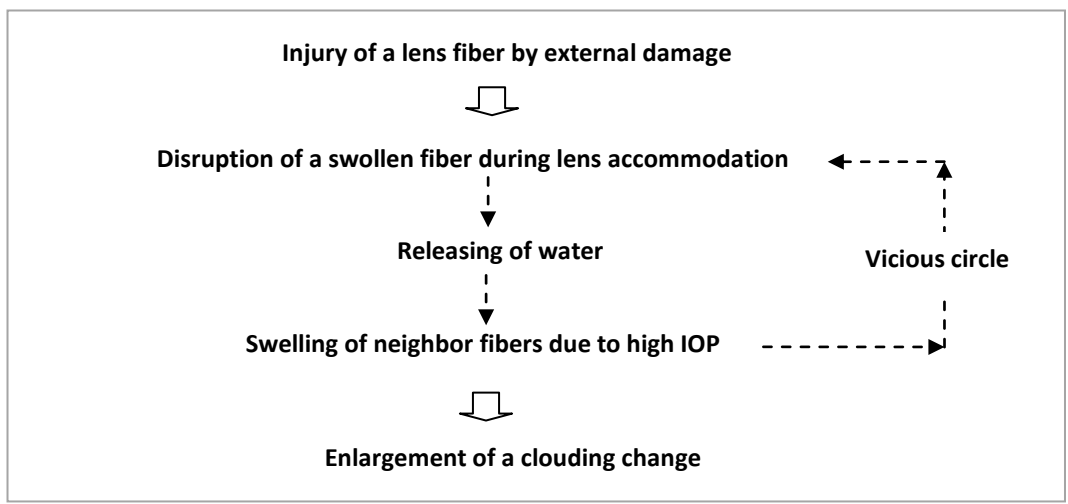

**Figure 2. A vicious circle between disruption of a swollen lens fiber and swelling of neighbor fibers**

Development of cortical cataract is promoted by injury of a cortical fiber by external damage. An injured fiber may disrupt when it is curved. Disruption of the fiber will release water. Due to high intracellular osmotic pressure (IOP), neighbor fibers can absorb water and become swelling. A swollen neighbor fiber may disrupt and release water again. Thus, a viscous circle is built up between disruption of a swollen fiber and swelling of neighbor fibers. This viscous circle drives successive swellings and disruptions of local fibers, and results in enlargement of a clouding change in cortical cataract.

In traumatic cortical cataract, a clouding change may also "grow" with time, but only in some patients. In our view, three factors are determinant for the consequence of a contusion-caused cataract: the location of injury, the degree of injury, and the age of patient. Contusion of the lens may cause disruption of a lens fiber and subsequent swelling of its neighbor fibers. However, when an injury is small and when it is in deep part of cortex, degeneration of lens cortex can be localized. Firstly, the amount of water that is released from a "dryer" fiber in deeper cortex is very limited, thus the degree of swelling of its neighbor fibers may be very low. Secondly, the fibers in deep cortex may have little load from the accommodations of the lens, thus the swollen fibers have low risk of disruption. Taken together, a traumatic cataract that occurs in the outer layers of lens cortex will have higher risk to "grow" than that in inner part of cortex. The age of patient may also affect the procession of a cataract. With age, accumulation of injuries makes lens fibers have increased fragility. A traumatic cataract that occurs to an old person may "grow" with time, but a similar cataract in a young person may stay unchanged for a long time.

### IV. The high IOP of a lens fiber as a driving force in development of cortical cataract

The clouding change in cortical cataract is progressive, and this is similar to that in aging changes, such as age spots and atherosclerotic plaques. However, the developing mechanism in cortical cataract is different from that in aging changes. In our view, aging of a living organism is a process of accumulation of Misrepairs of its structure (Wang-Michelitsch, 2015a). Here, the concept of Misrepair is defined as *incorrect reconstruction of an injured living structure.* Misrepair is a strategy of repair for maintaining the structural integrity for the survival of an organism in situations of severe injuries. Without Misrepairs, an individual could not survive to the age of reproduction; thus Misrepair mechanism is essential for the survival of a species. However, a Misrepair results in an irreversible alteration of a living structure, thus accumulation of Misrepairs disorganizes gradually the structure of a molecule, a cell, or a tissue, appearing as aging of it.

A Misrepair in a tissue leads to increased damage-sensitivity and reduced repair-efficiency of the local tissue. As a result, Misrepairs have a tendency to occur to the part of tissue where an old Misrepair has taken place. In return, new Misrepairs will increase again the damage-sensitivity of local tissue. By such a vicious circle, the frequency of Misrepairs to this part of tissue is increased and the range of affected tissue is enlarged after each time of Misrepair. Thus, accumulation of Misrepairs is focalized and self-accelerating (Wang-Michelitsch, 2015b). The focalized accumulation of Misrepairs results in development and enlargement of a "spot". Thus, the driving force in the "growing" of an aging change is the Misrepair itself.

Strictly speaking, development of cortical cataract is not a process of aging of the lens, since the body of lens cortex and lens nucleus is a non-living structure. However, in lens cortex, injuries and disruptions of lens fibers do not take place randomly. Due to the high IOP of a lens fiber, disruption of a fiber and swelling of its neighbor fibers compose a viscous circle, which results in the progressive degeneration of local lens cortex. Therefore, the high IOP of a lens fiber is the driving force in the development of cortical cataract.

**References**

*email : thomasjicun@gmail.com                                                                                                 7

*email : thomasjicun@gmail.com